\documentclass[aps,twocolumn,nofootinbib]{revtex4}
\usepackage[latin1]{inputenc}
\usepackage{epsfig}
\newcommand{\beq}{\begin{equation}}
\newcommand{\eeq}{\end{equation}}
\newcommand{\la}{\langle}
\newcommand{\ra}{\rangle}

\begin{document}

\title{Entropy production along a deterministic motion}

\author{Mário J. de Oliveira}
\affiliation{Universidade de São Paulo, Instituto de Física,
Rua do Matão, 1371, 05508-090 São Paulo, SP, Brazil}

\begin{abstract}

We propose a stochastic dynamics to be associated to a deterministic
motion defined by a set of first order differential equation. The
transitions that defined the stochastic dynamics are unidirectional
and the rates are equal to the absolute value of the velocity vector
field associate to the deterministic motion. From the stochastic
dynamics we determine the entropy production and the entropy flux. 
This last quantity is found to be the negative of the divergence of
the velocity vector field. In the case of a Hamiltonian dynamics
it vanishes identically.

\end{abstract}

\maketitle

%----------------------------------------------------------
\section{Introduction}

The calculation of the entropy $S$ of a system
by the Gibbs formula
\beq
S = - \int P(x)\ln P(x) dx,
\label{15}
\eeq
where $P(x)$ is the probability density distribution 
of the states $x$ of the system,
requires a description of the system by a probabilistic approach.
A probabilistic approach can be accomplished, for instance,
by considering that
the dynamic variables that characterize the state of 
the system are stochastic variables that obey a
stochastic equation of motion. Examples of this type
of equation are the master equation and the Fokker-Planck
equation for continuous space of states
\cite{kampen1981,tome2015L,tome2015}.

Let us consider a deterministic motion defined on a vector space
$x$ by the equation of motion
\beq
\frac{dx_i}{dt} = f_i(x),
\label{01}
\eeq
where $f_i$ are the components of a vector field $f(x)$
that depends on $x$, which we call the velocity vector field.
If a system is described by the deterministic dynamics
given by (\ref{01}) and
not by a stochastic dynamics, the question arises as to
how to determine entropy by the Gibbs formula.

An answer to the question just raised 
is to assign a probability distribution $P_0(x)$
at time $t=0$ and determine $P(x,t)$ for $t>0$ 
through (\ref{01}).
If $x=h(t;x')$ is the solution of (\ref{01}) with
the initial condition $x$ equal to $x'$ at $t=0$, then 
$P(x,t)dx=P_0(x')dx'$. From this equality, it is
straightforward to derive the equation
\beq
\frac{\partial P}{\partial t}
= - \sum_i \frac{\partial f_i P}{\partial x_i}
\label{02}
\eeq
by determining the Jacobian of the 
transformation defined by $x'\to x=h(t;x')$
where $t$ is understood as a parameter.
Equation (\ref{02}) was derived by Gerlich \cite{gerlich1973}
who called it the generalized Liouville equation.

Solving this equation for $P$,
we determine the entropy by the Gibbs formula.
Its time derivative follows from equation (\ref{02}) and is
\beq
\frac{dS}{dt} = \sum_i \la\frac{\partial f_i(x) }{\partial x_i} \ra,
\label{04}
\eeq
and expression derived by Dobbertin \cite{dobbertin1976}
and by Andrey \cite{andrey1985}. The right-hand side of (\ref{04})
is the average of the divergence of the velocity vector field $f$.

We may also determine from the equation (\ref{02}) the time
evolution of the average $s_i=\la x_i\ra$ which is given by
\beq
\frac{ds_i}{dt} = \la f_i(x)\ra.
\label{08}
\eeq

Let us suppose that the initial probability distribution is very
sharp at a certain point of the state space, that is, the 
initial state is a delta function. From equation
(\ref{02}) it follows that it remains forever sharp around the
average $s=\la x\ra$, and $\la f_i(x)\ra$ becomes $f_i(s)$,
and the equation (\ref{08}) reduces to
\beq
\frac{ds_i}{dt} = f_i(s),
\label{05}
\eeq
which is no other than equation (\ref{01}), if we
replace $s$ by $x$.
The sharpness of $P$ turns the equation (\ref{04}) into 
the equation
\beq
\frac{dS}{dt} = \sum_i \frac{\partial f_i(s)}{\partial s_i}
\label{06}
\eeq

We wish to show here that it is possible to set up another
stochastic dynamics that also fulfills the condition (\ref{05}),
and can thus be associated to the set of deterministic
equations (\ref{01}).
The basic assumption that we adopt is to interpret $|f_i(x)|=w_i$
as the rate of the transition $x_i\to x_i+\varepsilon$, 
where $\varepsilon$ is a small parameter that we set
to zero after we have calculated the
quantities of interest. The equation that we derive is
\beq
\frac{\partial P}{\partial t}
= - \sum_i \frac{\partial f_i P}{\partial x_i}
+ \frac{\varepsilon}2\sum_i \frac{\partial^2 w_i P}{\partial x_i^2},
\label{02a}
\eeq
which differs from equation (\ref{02}) by the additional
term proportional to $\varepsilon$.
For small enough $\varepsilon$, we reach (\ref{05}), as desired,
but $dS/dt$ is no longer given by (\ref{06}) and is given by
\beq
\frac{dS}{dt} = \sum_i \frac{\partial f_i(s)}{\partial s_i}
+ \frac12 \sum_i (\chi^{-1})_{ii} w_i,
\label{09}
\eeq
where $\chi$ is the solution of
\beq
\frac{d\chi_{ij}}{dt} = \sum_k (f_{ik}\chi_{kj} 
+ f_{jk}\chi_{ki}) + \delta_{ij} w_i.
\eeq
The negative of the first term in the right-hand side of
(\ref{09}) is a measure of the contraction of the 
volume of the state space, and was suggested 
to be the rate of entropy production 
\cite{gallavotti1995,ruelle1996,ruelle1997}.
Here, however we interpret it as the entropy flux, 
\beq
\Psi = -\sum_i \frac{\partial f_i(s)}{\partial s_i},
\eeq
that is, the flux of entropy from the system to the
outside. The second term in the right-hand side of
(\ref{09}) is nonnegative and we interpret it as
the rate of entropy production,
\beq
\Pi = \frac12 \sum_i (\chi^{-1})_{ii} w_i,
\label{03}
\eeq
which allows to write
\beq
\frac{dS}{dt} = \Pi-\Psi.
\eeq
that is, the time variation of the entropy is equal to the
rate of entropy production minus the flux of entropy to the
outside.

The stochastic dynamics that we consider here consists of
transitions that are unidirectional, also called absolute
irreversible transitions, and its entropy production has been
the subject of investigation in recent years 
\cite{ohkubo2009,murashita2014,zeraati2012,benavraham2011,%
busiello2020,rahav2014,saha2016,pal2017,pal2021,pal2021a,manzano2024}.
The entropy production associated to unidirectional transitions
cannot be calculated by the formula introduced by Schnakenberg
\cite{schnakenberg1976} because this formula requires the reverse
transition. The expression that we employ here for the entropy
production, which leads us to the formula (\ref{03}),
is a continuous version of a formula that we have
proposed for unidirectional transitions in discrete space
of states \cite{tome2024}.

%----------------------------------------------------------
\section{Stochastic motion}

\subsection{Master equation}

The stochastic motion is defined in such a way
that the vector
\beq
x=(x_1,x_2,\ldots,x_n)
\eeq
changes a distance $\varepsilon$ in
one of the $n$ direction of the state space.
The variable $x_i$ changes either to $x_i+\varepsilon$
if $f_i>0$ or to $x_i-\varepsilon$
if $f_i<0$. Defining $\sigma_i$ as the sign of
$f_i$ then $x_i$ changes to $x_i+\sigma_i\varepsilon$. 
The rate $w_i$ of this transition is chosen to 
be equal to the absolute value of $f_i$, that is,
$w_i(x)=|f_i(x)|$ and thus depends on $x$.

The equation that governs the time evolution of the
probability distribution $P(x)$ is then given by
\beq
\frac{dP(x)}{dt} = \frac1\varepsilon
\sum_i \{w_i(x-\sigma^i \varepsilon)P(x-\sigma^i \varepsilon),
- w_i(x)P(x)\}
\label{12}
\eeq
where $\sigma^i$ is the vector with all components equal to zero
except the $i$-th component which is $\sigma_i$.

The entropy $S$ is defined by the Gibbs formula
\beq
S = - \sum_x P \ln P,
\eeq
and its time derivative is given by
\beq
\frac{dS}{dt} = - \sum_x \frac{dP}{dt}\ln P dx.
\eeq
Using the master equation, we find
\beq
\frac{dS}{dt} = \frac1\varepsilon
\sum_i \sum_x w_i(x)P(x)\ln \frac{P(x)}{P(x+\sigma^i \varepsilon)}.
\eeq

The variation of entropy with time $dS/dt$ is due to the production
of entropy $\Pi$ minus the flux of entropy $\Psi$, that is,
\beq
\frac{dS}{dt} = \Pi - \Psi.
\eeq
The rate of entropy production is chosen to be 
given by the expression \cite{tome2024}
\[
\Pi= \frac1\varepsilon \sum_i \sum_x w_i(x)\times
\]
\beq
\times
\{ P(x)\ln \frac{P(x)}{P(x+\sigma^i \varepsilon)}
- P(x) + P(x+\sigma^i \varepsilon)\},
\eeq
which is nonnegative because each term is of the type
$\xi\ln\xi-(\xi-1)\geq0$.
The entropy flux is obtained from $\Psi=\Pi-dS/dt$,
that is, 
\beq
\Psi = - \frac1\varepsilon \sum_i \sum_x w_i(x)
\{P(x) - P(x+\sigma^i \varepsilon)\},
\eeq
which can be written as the average
\beq
\Psi = - \frac1\varepsilon \sum_i
\la\{w_i(x) - w_i(x-\sigma^i \varepsilon)\}\ra.
\label{28}
\eeq

The time evolution of $s_i=\la x_i\ra$ is given by
\beq
\frac{ds_i}{dt} = \la f_i(x)\ra,
\eeq
and we recall that $f_i=\sigma_i w_i$.

\subsection{Fokker-Planck equation}

To meet the condition (\ref{05}), we require the
probability $P(x)$ to be very sharp. This is
accomplished by considering that $\varepsilon$ is
small. 
Expanding the terms inside the curl brackets of equation 
(\ref{12}) up to second order in $\varepsilon$, we reach
the Fokker-Planck equation
\beq
\frac{\partial P}{\partial t}
= - \sum_i \frac{\partial f_i P}{\partial x_i} 
+ \frac\varepsilon2 \sum_i \frac{\partial^2 w_i P}{\partial x_i^2},
\label{13}
\eeq
where we used the relation $f_i=\sigma_i w_i$.

The entropy $S$ is defined by the Gibbs formula
\beq
S = - \int P \ln P dx,
\label{24}
\eeq
and its time derivative is
\beq
\frac{dS}{dt} = - \int \frac{\partial P}{\partial t}\ln P dx.
\eeq
Using the Fokker-Planck equation (\ref{13}),
and after an integration by parts, we find
\[
\frac{dS}{dt} = - \sum_i \int  f_i \frac{\partial P}{\partial x_i} dx
+ \frac\varepsilon2 \sum_i \int \frac{\partial w_i}{\partial x_i} 
\frac{\partial P}{\partial x_i} dx 
\]
\beq
+ \frac\varepsilon2 \sum_i \int \frac{w_i}{P}
(\frac{\partial P}{\partial x_i})^2. 
\label{14}
\eeq
We are assuming that $P(x)$ vanishes rapidly at the boundaries.

The rate of entropy production $\Pi$ is identified as 
the last summation in (\ref{14}),
\beq
\Pi = \frac\varepsilon2 \sum_i \int \frac{w_i}{P}
(\frac{\partial P}{\partial x_i})^2 dx,
\label{30}
\eeq
and is clearly nonnegative. The flux of entropy is then given by
\beq
\Psi = \sum_i \int  f_i \frac{\partial P}{\partial x_i} dx
- \frac\varepsilon2 \int \frac{\partial w_i}{\partial x_i} 
\frac{\partial P}{\partial x_i} dx.
\eeq
After an integration by parts, the flux of entropy can be
written as the average
\beq
\Psi = - \sum_i \la \frac{\partial f_i}{\partial x_i}\ra
+ \frac\varepsilon2
\sum_i \la \frac{\partial^2 w_i}{\partial x_i^2}\ra.
\label{34}
\eeq

The time evolution of $s_i=\la x_i\ra$ is also obtained
from the Fokker-Planck equation and reads
\beq
\frac{ds_i}{dt} = \la f_i\ra.
\label{32}
\eeq

\subsection{Gaussian distribution}

The second term in the right-hand side of (\ref{13})
is related to the fluctuations in $x$. Since it is
proportional to $\varepsilon$ we expect that the
fluctuations are small, which means that $P(x)$ is
very peaked at the average value of $x$. We then
expect that for sufficient small values of $\varepsilon$,
the average $\la f(x)\ra$ becomes $f(\la x\ra)$ and 
equation (\ref{32}) reduces to the equation (\ref{05})
as desired. We also expect that the expression (\ref{34})
for $\Psi$ reduces to
\beq
\Psi = - \sum_i f_{ii}(s)
\eeq
where $f_{ij}(x) =\partial f_i(x)/\partial x_j$.

To obtain these limiting results in a more precise procedure
we proceed as follows. We define the variables $y_i$ through
\beq
x_i = s_i + \sqrt\varepsilon\, y_i,
\eeq
where $s_i(t)$ depends on time and is the solution of
\beq
\frac{ds_i}{dt} = f_i(s).
\label{20}
\eeq
The probability density distribution $\rho(y)$
of $y$, a vector with components $y_i$, is related
to $P(x)$ by
\beq
\rho(y) = P(x)\varepsilon^{n/2}.
\eeq
From this equality we obtain
\beq
\frac{\partial \rho}{\partial t} = \varepsilon^{n/2}
\frac{\partial P}{\partial t}
+ \varepsilon^{n/2} \sum_i f_i(s)\frac{\partial P}{\partial x_i}.
\eeq
Replacing in this equation $\partial P/\partial t$ given by
the Fokker-Planck equation (\ref{13}), we reach the result
\beq
\frac{\partial\rho}{\partial t}
= \frac1{\sqrt\varepsilon}\sum_i
[f_i(s)\frac{\partial \rho}{\partial y_i}
- \frac{\partial f_i \rho}{\partial y_i}]
+ \frac12 \sum_i \frac{\partial^2 w_i \rho}{\partial y_i^2},
\eeq 
and we recall that in this equation,
$f_i$ and $w_i$ are functions of $x=s+\sqrt\varepsilon y$.

In the limit $\varepsilon\to0$, we find
\beq
\frac1{\sqrt\varepsilon}[f_i(s) - f_i(s+\varepsilon y)] \to 
\sum_j f_{ij}(s) y_j,
\eeq
and $w_i(s+\sqrt\varepsilon y)\to w_i(s)$.
Therefore, in the limit $\varepsilon\to0$, we reach the 
following equation for $\rho$,
\beq
\frac{\partial\rho}{\partial t}
= - \sum_{ij} \bar{f}_{ij} \frac{\partial y_j\rho}{\partial y_i}
+ \frac12 \sum_i \bar{w}_i \frac{\partial^2\rho}{\partial y_i^2},
\label{18}
\eeq 
where $\bar{f}_{ij} = f_{ij}(s)$ and $\bar{w}_i=w_i(s)$.

The solution of equation (\ref{18}) is the multivariate
Gaussian distribution
\beq
\rho(y) = \frac1Z \exp\{-\frac12 \sum_{ij}(\chi^{-1})_{ij},
y_i y_j\}
\label{22}
\eeq
where
\beq
Z = \int \exp\{-\frac12 \sum_{ij}(\chi^{-1})_{ij},
y_i y_j\} dy
\eeq
and the covariances $\chi_{ij}=\la y_iy_j\ra$ depend on $t$.
After calculating the integral
\beq
Z = (2\pi)^{n/2}[{\rm Det}(\chi)]^{1/2}.
\eeq

The equation that determines $\chi$ is obtained from the
Fokker-Planck equation (\ref{18}) and is
\beq
\frac{d\chi_{ij}}{dt}
= \sum_k (\bar{f}_{ik} \chi_{kj} + \bar{f}_{jk} \chi_{ki}).
+ \bar{w}_i \delta_{ij}
\label{21}
\eeq 
We remark that $s_i$ is identified as the average $s_i=\la x_i\ra$
and is the solution of (\ref{20}). After solving equation (\ref{20}),
we obtain $s(t)$ and replace this solution in $\bar{w}_i$
and $\bar{f}_{ij}$. The equation (\ref{21}) can then be solved for
$\chi$ and will depend on $t$.

To find the expression for $\Psi$ and $\Pi$ in the limit
$\varepsilon\to0$, we use the distribution $\rho$,
given by (\ref{22}). The entropy flux become
\beq
\Psi = - \sum_i \bar{f}_{ij}
= - \sum_i \frac{\partial\bar{f}_i} {\partial s_i},
\label{35}
\eeq
and the rate of entropy production become
\beq
\Pi = \frac12 \sum_i \int w_i \rho
(\frac{\partial \ln \rho }{\partial y_i})^2 dy.
\eeq
Using (\ref{22}) and performing the integral, we find the result
\beq
\Pi = \frac12 \sum_i \bar{w}_i (\chi^{-1})_{ii}.
\label{37}
\eeq

The entropy can be determined directly by the use of (\ref{22}) 
and it is
\beq
S =  \frac{n}2\ln\varepsilon +\frac{n}2  + \ln Z.
\eeq
Deriving $S$ with respect to $t$, we find
\beq
\frac{dS}{dt} = \frac12 \sum_{ij}(\chi^{-1})_{ij}
\frac{d\chi_{ij}}{dt}.
\eeq

Using the equation for $\chi_{ij}$ we get
\beq
\frac{dS}{dt} = \sum_i\bar{f}_{ii}
+ \frac12 \sum_i(\chi^{-1})_{ii}\bar{w}_i,
\eeq
and we see that the last term in the right-hand side is the rate
of entropy production and the first term is minus the entropy 
flux and this equation is consistent with $dS/dt=\Pi-\Psi$
as expected.

%----------------------------------------------------------
\section{Motion near a stable fixed point}

We consider here the motion near a stable fixed point
which we choose to occur at $x=0$. The equations of motion
around this point is chosen to be
\beq
\frac{dx_i}{dt} = \sum_j A_{ij} x_j.
\eeq
Performing a linear transformation $x\to x'$ so as to diagonalize
the matrix $A$ with elements $A_{ij}$, we reach the equations
\beq
\frac{dx_i'}{dt} = \lambda_i x_i',
\eeq
where $\lambda_i$ are the eigenvalues of $A$. The real parts of
the eigenvalues are negative or zero because we are assuming the
fixed point to be stable. 

We start our analysis by considering the equation of motion
associated to a variable $x_j'$ and that the corresponding
eigenvalue $\lambda_j$ is real. Using $x$ in the place of
$x_j'$, and defining $\lambda=|\lambda_j|$.  
\beq
\frac{dx}{dt} = - \lambda x.
\eeq
We suppose that the initial position is positive so that the
motion proceeds from right to left. In accordance with our
approach, the rate of the transition from $x$ to
$x-\varepsilon$ is $w(x)=\lambda x$ and the master 
equation reads
\beq
\frac{d}{dt}P(x) = \frac\lambda\varepsilon
\{(x+\varepsilon)P(x+\varepsilon)-xP(x)\}.
\label{25}
\eeq

From equation (\ref{25}), we obtain the time evolution
of the average $s=\la x\ra$,
\beq
\frac{ds}{dt} = - \lambda s,
\eeq
whose solution is
\beq
s = s_0 e^{-\lambda t},
\label{36}
\eeq
where $s_0$ is the value of $s$ at $t=0$.

Here we do not solve the master equation (\ref{25}) by
transforming it into a Fokker-Planck equation by an
expansion procedure, as we did above. Our procedure here
is to solve the equation (\ref{25}) directly as it is.

Defining $x=\varepsilon n$, the equation (\ref{25}) becomes
\beq
\frac{dp_n}{dt} = \lambda[(n+1)p_{n+1} - np_n],
\label{25a}
\eeq
valid for $n=0,1,2,\ldots$,
where $p_n = \varepsilon P(x)$.
We assume the solution of (\ref{25a}) to be a Poisson
distribution
\beq
p_n = \frac{e^{-\alpha }}{n!}  \alpha^n ,
\eeq
where $\alpha$ depends on $t$. Replacing it in equation
(\ref{25a}), we find the equation for $\alpha$,
\beq
\frac{d\alpha}{dt} = -\lambda \alpha,
\eeq
whose solution is
\beq
\alpha = \alpha_0 e^{-\lambda t}.
\eeq
From the Poisson distribution, $\la n\ra=\alpha$. 
Recalling that $x=\varepsilon n$, it follows that
$s=\la x\ra=\varepsilon \la n\ra = \varepsilon \alpha$
and $s_0=\varepsilon \alpha_0$.

The entropy is 
\beq
S = - \sum_n p(n)\ln p(n)
\eeq
which can be written as
\beq
S = - \alpha\ln \alpha + \alpha + \la \ln n!\ra,
\eeq
where
\beq
\la \ln n!\ra
= e^{-\alpha}\sum_{n=2}^\infty  \frac{\ln n!}{n!} \alpha^n.
\eeq

For $t$ sufficient large, $\alpha$ will be small and
retaining only the dominant terms in $S$ we find
\beq
S = - \alpha\ln \alpha + \alpha.
\eeq
The time derivative of $S$ is
\beq
\frac{dS}{dt} = \lambda\alpha\ln\alpha.
\eeq

The flux of entropy $\Psi$ is determined from (\ref{28})
and is simply $\Psi = \lambda$. 
Using $\Pi=dS/dt+\Psi$, the rate of entropy
production is
\beq
\Pi = \lambda + \lambda\alpha\ln\alpha.
\eeq
For sufficient large $t$, we see that $dS/dt$ vanishes and
$\Pi=\Psi=\lambda$.

%----------------------------------------------------------
\section{Conclusion}

We have proposed a stochastic dynamics to be associated to
deterministic motion defined defined by a set of first order differential
equation given by (\ref{01}). The associated stochastic
dynamics allowed us to determine not only the entropy
by also the rate of entropy production
and the flux of entropy, which are given by
the equations (\ref{37}) and (\ref{35}), respectively.

The entropy flux $\Psi$ is identified as the negative of
the divergence of the velocity vector field $f$,
\beq
\Psi = -\sum_i \frac{\partial f_i}{\partial x_i},
\eeq
where $f_i$ are the components of $f$.
If the dynamic system is a Hamiltonian system, it follows
immediately that the divergence of the vector
field vanishes and no flux of entropy is associated to
a Hamiltonian dynamics.
The rate of entropy production $\Pi$ is obtained from the
covariance matrix $\chi$, the solution of (\ref{21}),
and is a non negative quantity by construction.

If the system reaches a stationary state then
the time variation of the entropy $dS/dt$ vanishes
and $\Psi=\Pi$ and $\Psi \geq0$. This last result
is consistent with a result of Ruelle who showed
that the negative of the divergence of the velocity vector field
$f$ is nonnegative at the stationary state
\cite{ruelle1996,ruelle1997}. We remark however that
Ruelle called this quantity the production rate and 
not entropy flux as we did here.

%----------------------------------------------------------

\end{document}